\documentclass[twocolumn,showpacs,preprintnumbers,amsmath,amssymb]{revtex4}
\usepackage{graphicx}
\usepackage{dcolumn}
\usepackage{bm}
\usepackage{color}

\begin{document}


\title{A bolometric measurement of the antineutrino mass}

\author{ C.~Arnaboldi, C.~Brofferio, O.~Cremonesi, E.~Fiorini, C.~Lo\,Bianco, A.~Nucciotti,
M.~Pavan, G.~Pessina, S.~Pirro, E.~Previtali, M.~Sisti}
\author{L.~Martensson}
\altaffiliation{CEE fellow  in the Network on Cryogenic Detectors, under contract FMRXCT980167}
\affiliation{
Dipartimento di Fisica dell'Universit\`a di Milano-Bicocca \\
and Sezione di Milano dell'INFN, 
I-20126 Milano, Italy.}
\author{A.~Giuliani}
\affiliation{
Dipartimento di Scienze Chimiche, Fisiche e Matematiche  dell'Universit\`a \\ d'Insubria 
I-22100 Como, Italy 
and Sezione di Milano dell'INFN, 
 I-20133 Milano, Italy.
}
\author{B.~Margesin and M.~Zen}
\affiliation{
ITC-irst, Microsystems Division, I-38050 Povo (TN), Italy 
}
\

\begin{abstract}
High statistics calorimetric measurements of the $\beta$ spectrum of $^{187}$Re are being 
performed with arrays of silver perrhenate crystals operated at low temperature. 
After a modification of the experimental set-up, which allowed to substantially
reduce the background of spurious counts and therefore to increase
the sensitivity on the electron antineutrino mass, a new measurement with  10 silver
perrhenate microbolometers is  running since July 2002. The crystals
have masses between 250 and 350\,$\mu$g and their average FWHM energy resolution, 
constantly monitored by means of fluorescence X-rays, is of 28.3\,eV at the $\beta$ end-point.
The Kurie plot collected during 4485 hours$\times$mg effective running time has an
end-point energy of $2466.1 \pm 0.8_{stat} \pm 1.5_{syst}$\,eV, while the half lifetime of the
decay is found to be $43.2 \pm 0.2_{stat} \pm 0.1_{syst}$\,Gy. These values are the most precise
obtained so far for $^{187}$Re. From the fit of the Kurie plot 
we can deduce a value for the squared electron antineutrino mass $m_{\overline{\nu}_e}^2$ of
$147 \pm 237_{stat} \pm 90_{syst}$\,eV$^2$. The corresponding 90\%\,C.L. upper limit for 
$m_{\overline{\nu}_e}$ is 21.7\,eV. 

\end{abstract}

			      
\maketitle


The interest in direct measurements of  the neutrino mass  $m_\nu$  
from the $\beta$ decay
spectrum has been recently stimulated by  the evidence of a non zero value of
$\Delta m_\nu^{2}$ detected in searches on solar and atmospheric neutrino oscillations
 \cite{Fukuda01,Ahmad01,Nishikawa02}.
  
Limits on direct measurements of $m_\nu$ have been so far obtained
by experiments with electrostatic spectrometers investigating the $\beta$ decay of 
tritium. Recent results \cite{Bonn02,Part02}, 
have set an upper
limit  of 2.2\,eV at 90\% C.L.
These experiments are based on
the measurement of the spectrum of the emitted electron. One cannot therefore
exclude "a priori" that the  decay could partially occur on a excited state of the
daughter molecule. 
 This and other systematic effects have given in the past years a negative value
for m$_\nu$$^{2}$. Even if now these problems seem to be almost completely solved we think it 
is important to carry out a measurement in a different approach.

Calorimetric measurements where all the energy
released in the decay is recorded, appear therefore complementary to those carried 
out with  spectrometers. In addition they allow to measure the entire $\beta$ decay
spectrum, and can therefore test any possible distortion of the Kurie plot. 
A particularly suitable approach appears the bolometric one
\cite{Twerenbold96,Booth96} where detectors, operated at low temperature,  include absorbers 
of a material containing the $\beta$ active nucleus. 
If   the absorbers are diamagnetic and dielectric crystals their heat capacity can be 
very low, since  it is proportional to the cube of the operating temperature.
 As a consequence even the tiny energy delivered by a particle can give rise to a measurable  pulse
in a suitable thermal sensor. 
 
The present experiment is carried out on the first forbidden unique decay:

\begin{center}
$^{187}Re \to ^{187}Os + e^{-} + \overline\nu_{e}$ 
\end{center}

\noindent
which is particularly promising \cite{Vitale85} due to its low transition energy 
($\sim$2.5 keV). 
In addition the large isotopic abundance of
$^{187}$Re  (62.8\,\%)  allows the use of absorbers made with natural Rhenium. We
note also that  a precise direct measurement of the
 half lifetime of this decay 
 ($\sim$\,43\,Gy) is of great interest in
geochronology   for the determination of the age of minerals 
and meteorites  from their
Re-Os abundance \cite{Herr95,Reis95,Pearson99,Shir99,Alard02}.
Measurements of the spectra of
$^{187}$Re have been reported by the Genova group \cite{Galeazzi01}  with single crystals of
metallic Rhenium and by our group \cite{Alessandrello99} with an array of four crystals of 
  Silver Perrhenate (AgReO$_4$), a dielectric compound of Rhenium. 

In this paper we report on new high statistics measurements carried out 
in two successive runs with arrays of
ten crystals of AgReO$_4$ with masses ranging from 250 to 350\,$\mu$g.
 These crystals are thermally coupled to thermometers made of doped silicon chips 
 implemented by
the ITC-irst institute in Trento and tested and calibrated at low temperature in Milan 
\cite{Alessandrello99b}.  
Special care   is put  on the  
calibration of the energy scale 
and on the monitoring of the stability and performance of all detectors.
This is achieved by means of the 5.9 keV
K$\alpha$ line of $^{55}$Mn and the fluorescence K$\alpha$ lines produced by two $^{55}$Fe 
primary sources 
at 1.5, 2.6, 3.7 and 4.5 keV in Al, Cl, Ca, and   Ti, respectively. 
During the runs  all detectors are exposed to the fluorescence X-rays 
produced by the sources for $\sim$20 minutes  every two hours to allow a 
 continuous  energy calibration.   
In a first run  totaling   2354 hours$\times$mg  effective running time we have collected 
 $\sim 1.4 \times 10^{6}$ $^{187}$Re decays above an energy threshold of 
 700\,eV. 

The Kurie plot corresponding to the sum of all detectors was fit to
the spectrum calculated by W. Buhring  \cite{Buhring65}.  
The end-point energy for $^{187}$Re was found to be 
$2465.7 \pm 1.2 (stat.) \pm 1.5 (syst.)$\,eV. 
The systematic error is due to the uncertainties in the energy resolution, 
in the detector response function,
in the shape of the background below the beta spectrum and in the theoretical spectral 
shape for the $^{187}$Re beta decay. 
We have also attempted to determine the half lifetime of $^{187}$Re, by precisely
 measuring the mass of all absorbers and the total counts from each of them. 
\begin{figure}[t!]
\includegraphics[width=0.9\linewidth]{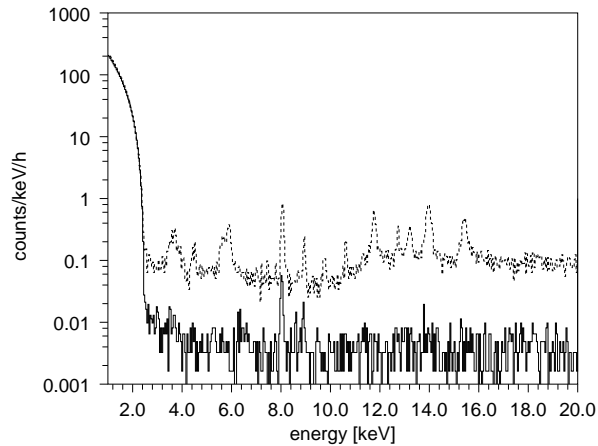}%
\caption{\label{fig:bkg} Background in the high energy region with (continuous line) 
and without (dashed line) source shielding. The peak at 8\,keV is the Cu K$\alpha$ line.}
\end{figure}
There were however spurious
counts which our Data Acquisition System could not eliminate.  
This  did not allow to determine precisely the  dead time of the measurement.
As a consequence we could only assume that all the running time was effective, 
thus determining  an upper limit of  45.1\,Gy  for the half lifetime of $^{187}$Re.  
No evidence was found for deviations 
from the calculated  Kurie plot. The squared electron antineutrino mass 
was found to be $335 \pm 499 (stat.) \pm 170 (syst.)$\,eV$^2$, where the systematic
error has the same origin as for the end-point energy quoted above.
A preliminary upper limit of 31.9\,eV at  90~\% C.L. 
could be  set on  the electron antineutrino mass. 
In this measurement the sensitivity  on the antineutrino mass is mainly 
limited by  the background of spurious counts, shown by the upper curve of Fig.\,\ref{fig:bkg}.
By means of a Monte Carlo simulation
 and of a dedicated measurement with the same set-up, but without the $^{55}$Fe primary
sources, we found out that 
  the background was mainly caused by the Internal Bremsstrahlung accompanying
the $^{55}$Fe E.C. decay, with a branching ratio of  $\sim3.2 \times 10 ^{-3}$\,\% \cite{Firestone98}. 
Most of the 
peaks appearing in this spectrum are due to fluorescence in the copper of the detector holders
and in a lead shield placed 
near the detectors.

\begin{figure}[t!]
\includegraphics[width=0.9\linewidth]{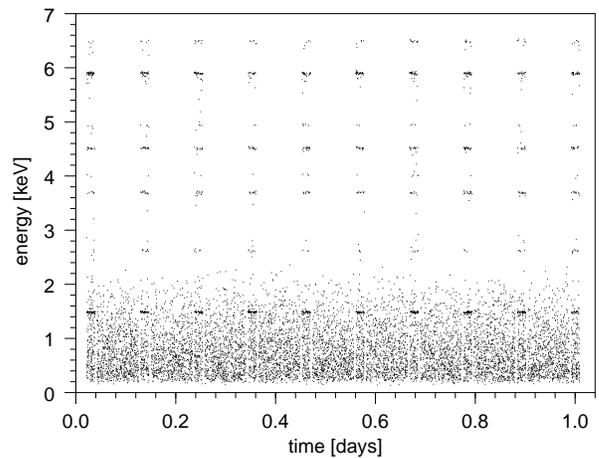}%
\caption{\label{fig:AvsT} A sample one day measurement showing the periodical exposure to the calibration source.}
\end{figure}

As a consequence we decided to substantially change the calibrating set-up with a 
new system which
automatically  moves  the sources in a  massive shield of Roman lead \cite{Alessandrello98}. 
The consequent reduction of the background is shown by 
 the lower curve of  Fig.\,\ref{fig:bkg}.
  The second run is being carried out in these improved conditions 
and with a new acquisition system which allows to precisely determine the live time
of the measurement. A  partially renewed  array of
10  AgReO$_4$ crystals with a total  mass of   2.683\,mg is running since July 2002.  
 The data from two detectors, with poor resolution,  are 
not included in our statistics: the corresponding effective total mass is therefore of 
2.174\,mg. 
 The present analysis refers to five months of continuous run totaling 4485 
and  1070  hours$\times$mg of effective   measurement   
 and calibration times, respectively.  
The FWHM  resolution of the  8 detectors  at 
 1.5\,keV (the Al K$\alpha$ line)  ranges from 21.2 to 28.7\,eV
with an average of  25.4\,eV, while the FWHM resolution of the
array extrapolated at the energy of the $\beta$ end-point (2.46\,keV) is 28.3\,eV. 
The  10 to 90\,\% risetime of the  detectors is in the  range 
340\,-\,680\,$\mu$s with an average value of 492\,$\mu$s.  
The check of the stability of the gain, which we consider essential for a correct 
measurement of the Kurie plot, 
is shown for a typical detector in Fig.\,\ref{fig:AvsT}.

\begin{figure}[t!]
\includegraphics[width=0.9\linewidth]{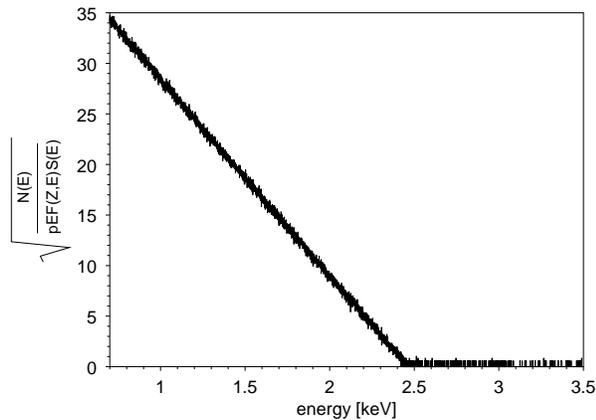}%
\caption{\label{fig:kurie} Kurie plot obtained, in the second run, from the sum of all 
8 detectors, where $p$ is the electron
momentum, $E$ is the electron kinetic energy, $F(Z,E)$ is the Coulomb factor and $S(E)$ 
is the shape factor.}
\end{figure}

The Kurie plot obtained from the sum of all  8 detectors is shown in Fig.\,\ref{fig:kurie}.
It corresponds to $\sim 3.2 \times 10^{6}$ $^{187}$Re decays above the common energy threshold of 
700\,eV. 
As in the preceding run, we find no deviation from the  above mentioned calculation of 
Buhring  for the spectrum expected in absence of the neutrino mass. The measured value for
the end-point is  $2466.1 \pm 0.8 (stat.) \pm 1.5 (syst.)$\,eV. 
The systematic error is again determined by the uncertainties in the energy resolution, 
in the detector response function,
in the shape of the background below the beta spectrum and in the theoretical spectral shape for 
the $^{187}$Re beta decay. 
Due to the substantial absence of spurious 
counts we could precisely determine the effective decay rate  from 
  the distribution of  the time 
intervals   between  two successive   $\beta$ decays. 
The half lifetime is thus found to be $[43.2 \pm 0.2(stat.) \pm 0.1(syst.)]\times 10^{9}$\,years, 
where the statistical error is mainly due to  the uncertainties in the measurement of the mass 
of the absorbers and the systematic error is due to the uncertainties in the pile-up 
discrimination. The values for the end-point energy and for the half lifetime are the most precise
existing in the literature. The latter, as noted before, has considerable impact in geocronology.

The squared electron antineutrino mass  
$m_{\overline{\nu}_e}^2$ is $147 \pm 237(stat.) \pm 90(syst.)$\,eV$^2$,
where the systematic
error has again the same origin as for the end-point energy quoted above.
The   90\,\% C.L. upper limit to   the electron antineutrino  mass is 21.7\,eV. 
This result is in agreement with the expected sensitivity  deduced from a Monte Carlo simulation of 
an experiment with the same statistical significance as our present data set. 
The limit on the electron antineutrino mass is not improved by a combined analysis of the two runs because of the
substantial improvements in the new run, both on the statistics and on the background. 

This result even if not yet competitive to those obtained with spectrometers 
has to be considered, 
in our opinion, complementary to them and show the potential of future 
bolometric measurements 
of the neutrino mass. 
As pointed out before it measures the  total released energy and not only the electron one, 
unless long living metastable states are formed. In addition it
allows  to investigate over the entire energy range all possible deviations from 
the standard  theory of  beta decay. 
The measurement is  still  presently running, but improvements are planned, based 
on the use of better and more massive crystals and  improved  thermal coupling between 
them and the thermistors. A new experiment is planned based on a larger array and different and 
faster thermal sensors. 

Thanks are due to  C.\,Callegaro, R.\,Cavallini, G.\,Ceruti, R.\,Gaigher,  
S.\,Parmeggiano,  M.\,Perego,   
and  to  our student   L.\,Soma   for continuous and constructive help  in
various  stages  of this experiment. We  also  gratefully acknowledge   contributions   of   
A.\,Alessandrello  and L.\,Zanotti in the first stage of this search. 
\\This  experiment  has  been  supported  in  part  by  the
Commission  of European Communities under contract  FMRX-CT98-0167.

\end{document}